\documentclass[%
reprint,
superscriptaddress,
twocolumn,
 amsmath,amssymb,
aps,
prb,
]{revtex4-2}

\usepackage{natbib}
\makeatletter
\def\NAT@spacechar{}
\makeatother

\usepackage{graphicx}
\usepackage{dcolumn}
\usepackage{bm}

\usepackage{xcolor}

\usepackage{caption}
\usepackage[hyperindex,breaklinks]{hyperref}
\usepackage{url}
\hypersetup{colorlinks=true,linkcolor=blue,citecolor=blue,urlcolor=blue}

\begin{document}

\newcommand{\changemarker}{\textcolor{black}}

\title{Training machine-learning potentials for crystal structure\\prediction using disordered structures}

\author{Changho Hong}
\thanks{These authors contributed equally to this paper.}
\author{Jeong Min Choi}
\thanks{These authors contributed equally to this paper.}
\author{Wonseok Jeong}
\thanks{These authors contributed equally to this paper.}
\author{Sungwoo Kang}
\thanks{These authors contributed equally to this paper.}
\author{Suyeon Ju}
\author{Kyeongpung Lee}
\author{Jisu Jung}
\affiliation{%
Department of Materials Science and Engineering and Research Institute of Advanced Materials, Seoul National University, Seoul 08826, Korea
}%
\author{Yong Youn}
\affiliation{%
Center for Green Research on Energy and Environmental Materials and International Center for Materials Nanoarchitectonics, National Institute for Materials Science, 1-1 Namiki, Tsukuba, Ibaraki 305-0044, Japan}%
\author{Seungwu Han}
 \email{hansw@snu.ac.kr}
 \affiliation{%
Department of Materials Science and Engineering and Research Institute of Advanced Materials, Seoul National University, Seoul 08826, Korea
}%

\date{\today}

\begin{abstract}
Prediction of the stable crystal structure for multinary (ternary or higher) compounds \changemarker{with unexplored compositions} demands fast and accurate evaluation of free energies in exploring the vast configurational space. The machine-learning potential such as the neural network potential (NNP) is poised to meet this requirement but a dearth of information on the crystal structure poses a challenge in choosing training sets. Herein we propose constructing the training set from density-functional-theory (DFT) based dynamical trajectories of liquid and quenched amorphous phases, which does not require any preceding information on material structures except for the chemical composition. To demonstrate suitability of the trained NNP in the crystal structure prediction, we compare NNP and DFT energies for Ba$_{2}$AgSi$_{3}$, Mg$_{2}$SiO$_{4}$, LiAlCl$_{4}$, and InTe$_{2}$O$_{5}$F over experimental phases as well as low-energy crystal structures that are generated theoretically. For every material, we find strong correlations between DFT and NNP energies, ensuring that the NNPs can properly rank energies among low-energy crystalline structures. We also find that the evolutionary search using the NNPs can identify low-energy metastable phases more efficiently than the DFT-based approach. By proposing a way to developing reliable machine-learning potentials for the crystal structure prediction, this work will pave the way to identifying unexplored multinary phases efficiently.\end{abstract}

\maketitle

\section{\label{sec1}Introduction}

Crystal structure prediction (CSP) for a given chemical composition is among the grand challenges in condensed matter physics~\cite{Maddox1988}. The goal of CSP is to identify atomic arrangements in space that produce the lowest free energy under given thermodynamic conditions. Mathematically, this is equivalent to the global optimization in a high-dimensional space, to which there is no general solution~\cite{Wolpert1997}. Nevertheless, various heuristic methods have been developed for navigating the gigantic configurational space efficiently and intelligently such as random structure sampling~\cite{Pickard2006,Pickard2011}, simulated annealing~\cite{Pannetier1990}, particle-swarm optimization~\cite{Wang2010,Call2007}, minima hopping~\cite{Goedecker2004}, basin hopping~\cite{Wales1997}, metadynamics~\cite{Marto2003}, and evolutionary algorithm~\cite{Oganov2006}. In evaluating the objective function or free energy, the method of choice is the first-principles calculations based on the density functional theory (DFT). The non-empirical nature of DFT allows for exploring the energy landscape with little restrictions, yet achieving high precisions. The DFT-based CSP has been successfully applied to identifying structures of organic crystals, superconducting materials, and inorganic crystals under extreme conditions~\cite{Zhu2012, Duan2015,OGANOV200695,OGANOV200838}.

One intriguing field to apply CSP is the structure prediction of ternary or higher (simply multinary hereafter) inorganic crystals at ambient conditions. Its importance arises from the low throughput of crystal synthesis: while structures of most unary and binary compounds were thoroughly investigated by X-ray crystallography, only about 16\% and 0.6\% have been revealed within the ternary and quaternary spaces, respectively~\cite{OGANOV201933}. In the Inorganic Crystal Structure Database (ICSD)~\cite{Bergerhoff1983} that collects most of the synthesized inorganic crystals, new entries accumulate at a pace of ~5,700 per year (averaged over year 2015--2019). Therefore, it may take at least several decades of strenuous experimental endeavors to uncover a large portion of ternary or quaternary domains in the structure database. Considering that major technological advances were achieved by multinary systems such as CuAlO$_2$~\cite{kawazoe1997p}, InGaZnO$_4$~\cite{Nomura1269}, CuIn$_x$Ga$_{1-x}$Se$_2$~\cite{ doi:10.1002/pip.822}, and Li$_{10}$GeP$_2$S$_{12}$~\cite{kamaya2011lithium}, a rapid knowledge expansion on multinary compounds via efficient and reliable CSP, followed by theoretical investigations on their properties, may expedite discovery of novel materials from the uncharted chemical space.

 There have been several theoretical attempts to identify the stable structures of unknown ternary systems\changemarker{~\cite{PhysRevMaterials.3.035404,doi:10.1021/acs.inorgchem.7b02102,PhysRevMaterials.2.023604,PhysRevMaterials.1.074803,doi:10.1021/jp100990b,shao2019ternary,GUBAEV2019148,shi2017high,doi:10.1021/cm100795d,doi:10.1021/jacs.7b08460,PAVLOUDIS201622,doi:10.1021/cm051601y,saleh2016novel,qu2020high,doi:10.1021/jacs.7b04456,doi:10.1021/acs.chemmater.7b00809,doi:10.1021/acs.chemmater.5b00716,wang2019discovery,OGANOV200695,OGANOV200838,doi:10.1063/5.0015672}}. However, the previous works often limited the configurational space by referring to available prototypes or fixing the number of atoms in the unit cell. The main reason for the restrictive searching is the sheer computational cost of DFT calculations. To predict the equilibrium structure of multinary crystals from exhaustive searching, \changemarker{a significant} increase of possible atomic arrangements \changemarker{from permutations among different species} demands evaluating the objective function far more efficiently than DFT. The classical potential is computationally cheap, which allowed for predicting stable structures of even quinary compounds~\cite{collins2017accelerated}. However, scarcity and low fidelity of classical potentials prohibit wide use as an objective function. Recently, machine-learning potentials (MLPs) are drawing much attention as it delivers accuracy of DFT yet is faster than DFT by more than thousands fold. (Exploiting the locality of quantum systems, the MLP is a variant of the order-\textit{N} method for DFT calculations, which is realized by atomic energies~\cite{PhysRevMaterials.3.093802}.) Therefore, MLP is poised to meet the requirement for evaluating energies in multinary CSP. Indeed, there are recent works that pursued this direction~\cite{PhysRevLett120156001,PhysRevB99064114,C8FD00055G,cphc.201700151}

The development style of MLP invokes a distinct challenge when applying MLP to CSP for unknown multinary compounds. That is to say, MLP infers total energies after learning on DFT results for reference structures. In usual practices, MLPs are first trained over structures derived from known crystals. However, such information is not available at the outset in CSP for unknown compounds, and one should construct MLPs out of `nothing'.
To cope with this hurdle, in Refs.~\cite{PhysRevLett120156001,PhysRevB99064114,C8FD00055G},  structures from the random search or evolutionary algorithm were used to generate and update MLPs for unary systems of B, C, and Na, whenever prediction uncertainties are large for the structures (meaning that they are yet to be learned). However, these approaches might be less effective for multinary compounds \changemarker{because they unnecessarily sample high-energy structures that violate chemical rules from mutation (genetic algorithm) or random distribution (random structure search).}

Motivated by the above discussions, we herein propose a way to construct MLP as a hi-fidelity surrogate model of DFT, mainly targeting to predict the most stable structure of multinary inorganic crystals at ambient conditions. The key strategy is to train an MLP over disordered structures such as liquid and melt-quenched amorphous phases. With only compositional information, the molecular-dynamics (MD) simulation on liquids can self-start from a random distribution and quickly equilibrates at sufficiently high temperatures (well above melting points), which is then cooled to amorphous structures. Thus, it is feasible to build the whole training set without preceding knowledge on the crystal structure. Furthermore, short-range orders in the amorphous phase resemble those in the crystalline phase (for example, consider amorphous Si and SiO$_2$), and local fluctuations in liquid and amorphous phases also sample diverse local orders that can appear in crystals. Therefore, it is anticipated that the trained MLP provides correct energies for stable as well as metastable phases, thus properly ranking energies of various structures emerging from search algorithms. We note that in Ref.~\cite{cphc.201700151}, amorphous and liquid structures of C were used in training a MLP which could in turn identify metastable phases for C. However, the application to finding the crystal structure of multinary systems has not been attempted as far as we are aware. As a machine-learning model, we adopt the Behler-Parrinello-type neural-network potential (NNP)~\cite{PhysRevLett.98.146401}. However, the present method is also compatible with other MLPs. For test materials, we select Ba$_2$AgSi$_3$, Mg$_2$SiO$_4$, LiAlCl$_4$, and InTe$_2$O$_5$F whose experimental structures are available at ICSD. These compounds also encompass diverse chemical bonding and structural motifs. 

The main objective of this article is to benchmark NNPs trained over disordered structures as a surrogate model of DFT in structure searching. To this end, we compare energies between DFT and NNP for stable and metastable structures. A strong correlation may indicate that the trained NNP can properly rank diverse structures explored by the CSP algorithm. To collect a large pool of metastable structures, we utilize an evolutionary algorithm in combination with DFT calculations, which is a favored approach in CSP~\cite{Oganov2006}. Then, we compare NNP and DFT energies for each compound over experimental phases in ICSD as well as low-energy metastable structures that emerge from the evolutionary algorithm. For every material, we find strong correlations between DFT and NNP energies, confirming that the NNPs can accurately prioritize structures in the order of the total energy. Interestingly, NNPs consistently predict that an experimental structure is the most stable. By combining the evolutionary algorithm with the trained NNP, we also search the stable structure for the test compounds.  The method finds the stable phase for LiAlCl$_4$ and metastable structures for other compounds with energies higher than for the stable phases by 11.5-41.2 meV/atom, and they are obtained more efficiently compared to the DFT-based approach. The rest of the paper is organized as follows: in Sec.~\ref{sec2} we explain computational methods used in the present work. The main results are discussed in Sec.~\ref{sec3} and Sec.~\ref{sec4} summarizes and concludes this work.   

\section{\label{sec2}COMPUTATIONAL METHODS}
\subsection{\label{sec2A}Construction of training set}
To generate training structures for NNP to be used in CSP, we carry out first-principles molecular dynamics (FPMD) simulations on the melt-quench-annealing process for each material. All DFT calculations in the present work are performed with Vienna \textit{ab initio} simulation package (VASP)~\cite{KRESSE199615} and the Perdew-Burke-Ernzerhof (PBE) functional is used for the exchange-correlation energy of electrons~\cite{PhysRevLett.77.3865}. The cutoff energies and the \textbf{k}-point meshes for FPMD simulation are determined by the convergence test on a \changemarker{superheated structure at 4000 K (see below)} such that the energy, pressure, and the maximum atomic force converge within 20 meV/atom, 10 kbar, and 0.3 eV/{\AA}, respectively. \changemarker{As results, 250, 450, 300, and 500 eV of cutoff energies are chosen for Ba$_2$AgSi$_3$, Mg$_2$SiO$_4$, LiAlCl$_4$, and InTe$_2$O$_5$F, respectively, and the $\Gamma$ point is used in the Brillouin zone integration for all the materials. }

We first determine the melting temperature (\textit{T}$_{\rm m}$) and simulation volume as follows: the initial structure is prepared by randomly distributing $\sim$100 atoms for the given stoichiometry, which is then superheated at 4000 K for 5 ps. Next, we perform FPMD simulations by lowering the temperature gradually and select an ad hoc \textit{T}$_{\rm m}$ as the lowest temperature at which the mean square displacement of atoms linearly increases with time. The determined \textit{T}$_{\rm m}$'s are 1500, 3500, 1500, and 2000 K for Ba$_2$AgSi$_3$, Mg$_2$SiO$_4$, LiAlCl$_4$, and InTe$_2$O$_5$F, respectively. (The experimental \textit{T}$_{\rm m}$'s are 2174~\cite{doi:10.1029/93GL01836} and 419 K~\cite{Weppner_1977}  for Mg$_2$SiO$_4$ and LiAlCl$_4$, respectively.) The cell volume is then adjusted such that the average hydrostatic pressure is equal to zero. Using the obtained \textit{T}$_{\rm m}$ and cell parameters, we generate liquid-phase trajectories for 20 ps in the NVT condition. Subsequently, the liquid is quenched with a cooling rate of 100 K/ps from \textit{T}$_{\rm m}$ to 300 K, and then annealed at 500 K for 15 ps to sample amorphous structures. The training set samples the whole melt-quench-annealing trajectory every 20 fs, and consists of 2,400-3,200 \changemarker{structures}. 

\changemarker{In the above, the computational parameters were determined by testing convergence for superheated liquid structures. When tested over 30 snapshots sampled from the training set of each material, the same computational settings result in mean absolute errors below 4 meV/atom and 0.12 eV/{\AA} for the total energy and atomic forces, respectively, which provides sufficient accuracy as training data. (The reference data are obtained with $3\times 3 \times3$ {\bf k}-point sampling and the cutoff energy of 700 eV.)
}

\subsection{\label{sec2B}Training NNP}
For training NNPs, we use the SIMPLE-NN package~\cite{LEE201995} that utilizes Google Tensorflow for training and LAMMPS~\cite{PLIMPTON19951} for MD simulations. The package has been applied to studying amorphous phases of GeTe~\cite{LEE2020109725} and Ni-silicidation process~\cite{doi:10.1021/acs.jpclett.0c01614}. \changemarker{As an input descriptor, we employ Behler-Parrinello type atom-centered symmetry function vectors (\textbf{G}), $\textit{G}^\textup{2}$  and $\textit{G}^\textup{4}$ for radial and angular distributions, respectively, following the definition in Ref.~\cite{doi:10.1063/1.3553717}. For each pair of atomic species, 8 radial and 18 angular components are used with the cutoff radius of 6 {\AA}, resulting in 132 input layers for ternary materials and 212 for quaternary material.} The network architectures are 132-30-30-1 and 212-30-30-1 for ternary and quaternary compounds, respectively. The output layer provides an atomic energy [\textit{E}$_\textup{at}($\textbf{G}$_\textit{i}$)] for an atom $i$ in the given structure, and the atomic energies add up to the total energy (\textit{E}$_\textup{tot}$) of the structure. The input vectors are decorrelated by principal component analysis (PCA), and then whitened, which significantly increases the learning speed~\cite{LEE2020109725}. We train NNPs by minimizing the loss function that is the sum of \changemarker{mean-square errors} for energy, force, and stress. The weight parameters in NNP are updated in a minibatch style with the momentum-based Adam optimizer~\cite{kingma2014adam}. We also include an L2 regularization term in the loss function to avoid overfitting and allocate 10\% of the training set to the validation set. The optimization is performed until the \changemarker{root-mean-square-errors} of the validation set become smaller than 10 meV/atom, 0.2 eV/{\AA} \changemarker{(0.4 eV/{\AA} for InTe$_2$O$_5$F),} and 10 kbar for energy, force, and stress components, respectively. \changemarker{ We also confirm that the number of reference structures in the training set is sufficient by checking the convergence of the validation error with respect to the training set size (not shown).}

\subsection{\label{sec2C}Crystal structure prediction by genetic algorithm}
To sample metastable structures for each material, CSP is performed by an evolutionary algorithm implemented in the Universal Structure Predictor: Evolutionary Xtallography (USPEX) package~\cite{GLASS2006713} while the energy evaluation is carried out by DFT calculations with VASP. We fix the number of formula units (\textit{Z}) to that of the stable structure (4 for every material). We set the population size to 20-60, which increases with the number of atoms in the unit cell. Initial structures are generated by either random symmetric~\cite{10.2138/rmg.2010.71.13} or topological structure generators~\cite{BUSHLANOV20191}. The succeeding structures are produced by both random generators and evolutionary operators, including heredity, permutation, soft mutation, and lattice mutation. The ratio of variation operators in USPEX is set automatically, encouraging the operators to produce more diverse structures in the low-energy spectrum~\cite{BUSHLANOV20191}. The generated structures are fully relaxed (both atomic positions and lattice vectors) until atomic forces and total stress become less than 0.1 eV/{\AA} and 20 kbar, respectively, or the number of relaxation steps reaches 400. In addition, we turn on the antiseed option, which prevents the evolution from being trapped in local minima by adding repulsive Gaussian potentials for sampled structures~\cite{LYAKHOV20131172}. We collect metastable structures generated during the whole evolution and use them in benchmarking NNPs. For accurate evaluation of energies, the metastable structures are further relaxed by DFT calculations until atomic forces and stresses are less than 0.02 eV/{\AA} and 4 kbar. USPEX is also used in predicting stable structures within NNP by interfacing with LAMMPS. (See Sec.~\ref{sec3C}.) 

\section{\label{sec3}RESULTS AND DISCUSSIONS}
\subsection{\label{sec3A}Test materials}
\begin{figure}
\centering
 \noindent\makebox[\linewidth]{\includegraphics[scale=0.8]{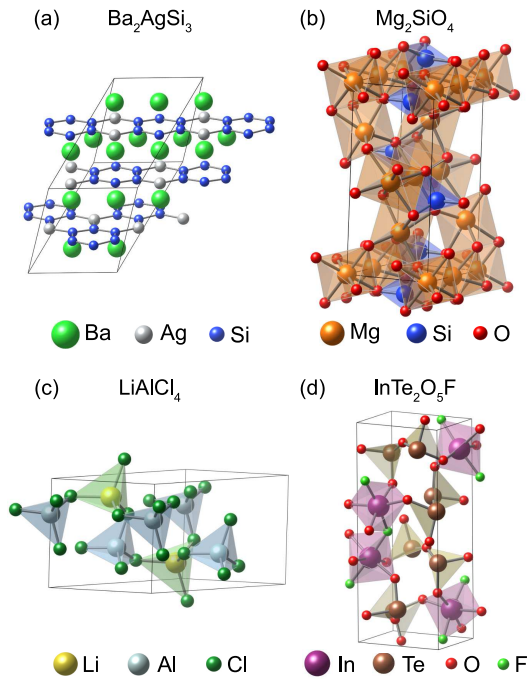}}
\caption{(Color online) Unit cells of the stable phase of (a) Ba$_2$AgSi$_3$, (b) Mg$_2$SiO$_4$, (c) LiAlCl$_4$, and (d) InTe$_2$O$_5$F \label{fig1}}
\end{figure}
 For test materials, we choose four compounds (three ternary and one quaternary) from ICSD: Ba$_2$AgSi$_3$ (\textit{Fddd}), Mg$_2$SiO$_4$ (\textit{Pnma}), LiAlCl$_4$ ({\textit{Pmn}2$_1$) and InTe$_2$O$_5$F (\textit{C}222$_1$). (See Fig.~\ref{fig1}.) These four materials encompass diverse structural motifs such as layers, intercalations, and shared polyhedra. \changemarker{We intentionally chose materials with low symmetries (rather than simple systems with the high symmetry such as SrTiO$_3$) to stress-test the proposed approach.} These materials are employed in a variety of applications owing to their mechanical, electrical, and optical properties: Ba$_2$AgSi$_3$ is a member of the Ba-Ag-Si system, which is anticipated for potential high-\textit{T}$_{\rm c}$ superconductors~\cite{HERRMANN199829}, and has a layered structure formed by Si$_6$~\cite{Gil1999}. Mg$_2$SiO$_4$ (also known as forsterite) features high fracture toughness and is actively studied as bioceramic implants~\cite{NI200783}. It has a shared-polyhedra structure with Mg and Si occupying octahedral and tetrahedral sites, respectively. There are also three additional structures of Mg$_2$SiO$_4$ available in ICSD with the space group of \textit{Ibmm}, \textit{Fd}$\overline{3}$\textit{m}, and \textit{Cmc}2$_1$. LiAlCl$_4$ is an archetypal halide solid-state Li-ion conductor~\cite{WEPPNER1976245}. After aliovalent doping, LiAlCl$_4$ exhibits high Li-ion conductivities, which could be used as electrolytes in all-solid-state batteries~\cite{Jeong_submit,he2017origin}. There are two structures of LiAlCl$_4$ in ICSD with the space group of \textit{P}2$_1$/\textit{c} and \textit{Pmn}2$_1$. Among them, we find that the \textit{Pmn}2$_1$ structure has a slightly lower DFT energy at 0 K~\cite{Promper:wm5410}. LiAlCl$_4$ has a relatively simple structure where both Li and Al ions are tetrahedrally coordinated by Cl ions. Lastly, InTe$_2$O$_5$F is anticipated for nonlinear optical applications owing to a non-centrosymmetric structure~\cite{JenneneBoukharrata:fn3131}. The In ions occupy octahedral sites surrounded by four O and two F ions. \changemarker{The theoretical band gaps calculated within the hybrid functional~\cite{doi:10.1063/1.1564060} are  0.25, 6.70, 7.17, and 4.71 eV for Ba$_2$AgSi$_3$, Mg$_2$SiO$_4$, LiAlCl$_4$, and InTe$_2$O$_5$F respectively.} 
 
\subsection{\label{sec3B}Comparison of NNP and DFT energies for metastable structures}
To judge the suitability of the developed NNPs for multinary CSP, we compare NNP and DFT energies for metastable structures collected by USPEX (see Sec.~\ref{sec2C}) as well as experimental structures in ICSD (relaxed within PBE). During 10-20 generations, USPEX garnered on average ~274 metastable structures with energies relative to that of the stable phase ($\Delta$\textit{E}$_\textup{tot}$) less than 500 meV/atom. The lowest $\Delta$\textit{E}$_\textup{tot}$'s are 46.5, 28.2, 1.9, and 33.2 meV/atom for Ba$_2$AgSi$_3$, Mg$_2$SiO$_4$, LiAlCl$_4$, and InTe$_2$O$_5$F, respectively. (The package could not identify experimental structures.) We note that many metastable structures share similar atomic configurations such that slight shifts of a few atoms relax one metastable structure to another. 

\begin{figure}
\centering
 \noindent\makebox[\linewidth]{\includegraphics[scale=0.95]{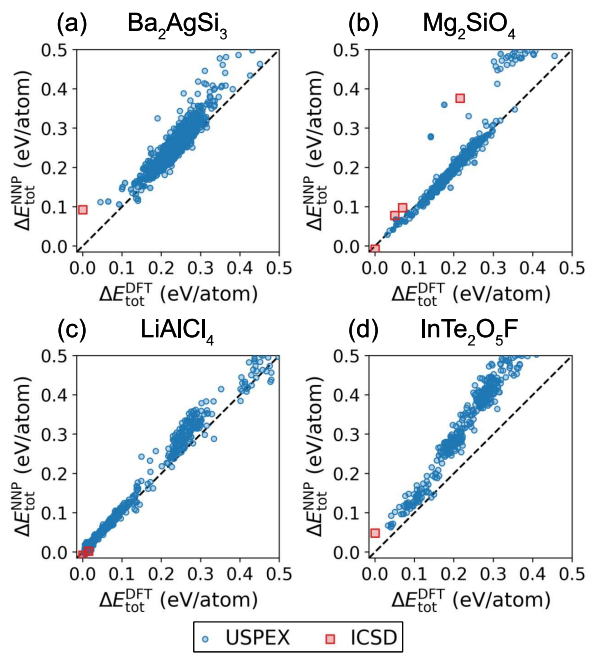}}
\caption{(Color online) Correlation between DFT and NNP energies. Structures are fixed to metastable structures from USPEX (blue circles) or experimental structures from ICSD that are relaxed by DFT (red squares).  In both $\Delta$\textit{E}$_\textup{tot}^\textup{NNP}$ and $\Delta$\textit{E}$_\textup{tot}^\textup{DFT}$, the reference energy is the DFT energy of the stable phase. The experimental structures are plotted as red squares. (a) Ba$_2$AgSi$_3$, (b) Mg$_2$SiO$_4$, (c) LiAlCl$_4$, and (d) InTe$_2$O$_5$F.  \label{fig2}}
\end{figure}

  Using NNPs trained over disordered structures (see Sec.~\ref{sec2B}), we evaluate energies for metastable structures without further relaxations and compare them with DFT energies in Fig.~\ref{fig2}. It is seen that the NNP and DFT energies are highly correlated, and Pearson coefficients among the structures with $\Delta$\textit{E}$_\textup{tot}^\textup{DFT}$ $\textless$ 200 meV/atom are 0.769, 0.864, 0.977, and 0.962 for Ba$_2$AgSi$_3$, Mg$_2$SiO$_4$, LiAlCl$_4$, and InTe$_2$O$_5$F, respectively. This is striking because none of the metastable structures were explicitly included in the training set. Thus, it is confirmed that the training set could sample the structural motifs appearing in low-energy metastable phases. It is also intriguing in Fig.~\ref{fig2} that the NNPs consistently predict the most stable experimental structure to be more stable than any theoretical structures (see red squares). This is the case even if the structures are relaxed using the NNPs.

  Each disordered structure in the training set consists of various local configurations that are similar to structural motifs in metastable phases. The strong energy correlations in Fig.~\ref{fig2} suggest that the machine-learning procedure successfully delineated local energies without ad hoc energy mapping~\cite{PhysRevMaterials.3.093802}.

\changemarker{ In Fig.~\ref{fig2}, RMSEs of NNP energies with respect to DFT energies are 27.4, 29.6, 10.7, and 63.7 meV/atom for Ba$_2$AgSi$_3$, Mg$_2$SiO$_4$, LiAlCl$_4$, and InTe$_2$O$_5$F, respectively (averaged over structures within 200 meV/atom). These are bigger than training errors, which is understandable because local orders in disordered structures do not exactly match with those of crystalline phases. Nevertheless, the trained NNPs can still serve as a surrogate model because candidate structures are recalculated by DFT (see the next subsection).}

In Fig.~\ref{fig2}, it is also seen that the errors of NNP prediction increase with $\Delta$\textit{E}$_\textup{tot}$. This \changemarker{might be} because structural features in high-energy phases \changemarker{({\textgreater} 200 meV/atom)} were not sufficiently sampled in the \changemarker{entropy-driven} liquid or amorphous structures. ($\Delta$\textit{E}$_\textup{tot}$ of the amorphous structure is 60--180 meV/atom.) \changemarker{The systematic upward deviation can be explained because structures are relaxed by DFT, which slightly deviate from the equilibrium positions given by NNP. When the structures are relaxed by NNP, the same plots show downward deviations (not shown).}

  In Fig.~\hyperref[fig2]{2(a)}, we note that the energy scale is not well resolved for low-energy structures of Ba$_2$AgSi$_3$. For instance, $\Delta$\textit{E}$_\textup{tot}$ for the third lowest metastable structure is 93 meV/atom in DFT but it is only 14 meV/atom in NNP. This is attributed to deficiencies in the training set: in the disordered phases, we find that the hexagonal Si ring in the crystalline structure [see Fig.~\hyperref[fig1]{1(a)}] is absent and linear Si chains embedded with Ag atoms are prevalent. Since the low-energy structures mainly differ in the connection topology of Si chains, the energy prediction in this region becomes rather inaccurate. This sampling problem is caused by the high cooling rate of 100 K/ps, which may not provide enough time to establish medium-range orders such as hexagonal rings. 
    
In Fig.~\ref{fig3}, using the principal component analysis, we examine element-by-element distributions of \textbf{G} vectors in the training set and (meta)stable structures of Ba$_2$AgSi$_3$. It is seen that most \textbf{G} points from USPEX-generated and experimental structures lie within those from the training set. (Analysis on other principal axes show similar relationship\changemarker{s}.) Other materials also exhibit \textbf{G}-point distributions similar to Ba$_2$AgSi$_3$. (The only exception is Al in LiAlCl$_4$, where Al atoms from high-energy metastable structures are distinct from the training set.) Therefore, we explicitly confirm that the local motifs of USPEX-generated structures and ICSD structures are well-included in the region spanned by the training set.
 
 \begin{figure}
\centering
\noindent\makebox[\linewidth]{\includegraphics{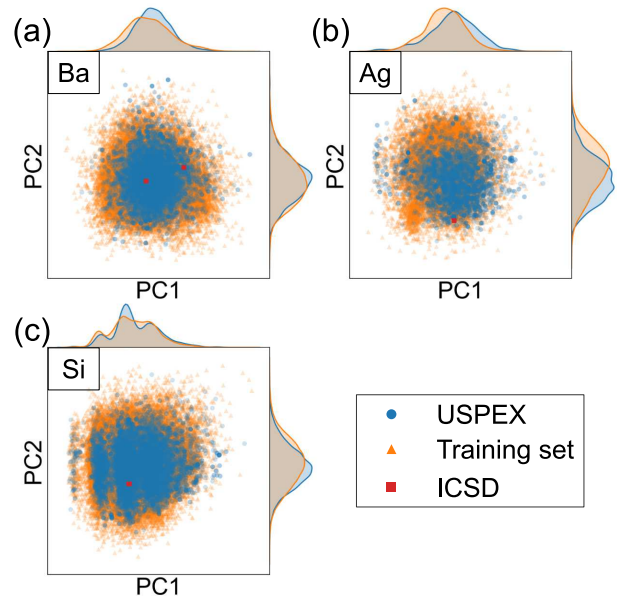}}
\caption{(Color online)  The  distribution of \textbf{G} vectors in Ba$_2$AgSi$_3$  projected onto the first two principal component axes (PC1 and PC2). The distribution of (a) Ba atoms, (b) Ag atoms, and (c) Si atoms. The projected density on each axis is plotted on the top and side.} \label{fig3}
\end{figure}

\subsection{\label{sec3C}Prediction of crystal structures combining USPEX and NNP}
Given that the NNPs trained over disordered phases are suitable as a hi-fidelity model in CSP, we try to search the stable phase for the test compounds by interfacing USPEX with LAMMPS and SIMPLE-NN. The computational parameters in USPEX are the same as those in Sec.~\ref{sec2C} except that the evolution extends to 120 generations. This is much longer than 10-20 generations in the DFT-based approach (see Sec.~\ref{sec2C}) but the total computational time is shorter owing to the fast evaluation by NNP. After 120 generations, USPEX suggests 10-20 candidate structures with lowest energies. By utilizing AMP$^2$~\cite{YOUN2020107450}, an automation script for operating VASP, we relax these structures until all the atomic forces and stress components are less than 0.02 eV/{\AA} and 4 kbar, thereby obtaining accurate DFT energies.

\begin{figure}
\centering
 \noindent\makebox[\linewidth]{\includegraphics[scale=0.8]{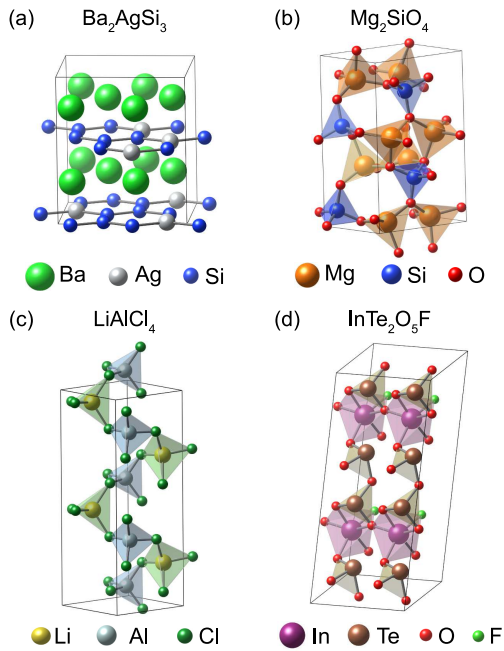}}
\caption{(Color online) The lowest-energy structures of (a) Ba$_2$AgSi$_3$, (b) Mg$_2$SiO$_4$, (c) LiAlCl$_4$, and (d) InTe$_2$O$_5$F found by USPEX in combination with NNP. \label{fig4} (c) is the same as Fig.~\hyperref[fig1]{1(c)}.}
\end{figure}

Despite a larger number of generations compared to previous studies (mostly binary compounds)~\cite{LYAKHOV20131172,LYAKHOV20101623,doi:10.1021/acs.jpcc.9b03274}, the evolutionary algorithm identifies the most stable phase only for LiAlCl$_4$ and fails for the other three materials. This may indirectly reflect complicated configuration spaces for multinary compounds. Figure~\ref{fig5} shows the structures with lowest $\Delta$\textit{E}$_\textup{tot}$ (41.2, 27.7, 0, and 11.5 meV/atom for Ba$_2$AgSi$_3$, Mg$_2$SiO$_4$, LiAlCl$_4$, and InTe$_2$O$_5$F, respectively). While local structures of three metastable phases look similar to those in the stable phase  (see Fig.~\ref{fig1}), there are also distinct differences; Ba$_2$AgSi$_3$ in Fig.~\hyperref[fig4]{4(a)} consists of only mixed Si-Ag rings without pure Si rings in Fig.~\hyperref[fig1]{1(a)}. On the other hand, Mg$_2$SiO$_4$ in Fig.~\hyperref[fig4]{4(b)} comprises corner-sharing tetrahedral Mg and Si atoms while the stable phase in Fig.~\hyperref[fig1]{1(b)} is characterized by edge-sharing octahedral Mg and tetrahedral Si atoms. (The corner-sharing structures were also obtained by DFT-based USPEX in the previous subsection.) Nevertheless, the NNP energy of the stable phase is close to that of DFT, which implies that more aggressive mutations are in need to create and stabilize the octahedral configuration of Mg during evolution. In the case of InTe$_2$O$_5$F, although both the experimental and theoretical structures have three-fold Te and two-fold O, substantial differences are observed for In octahedra. The octahedra in Fig.~\hyperref[fig1]{1(d)} are corner-shared with each other through F atoms, resulting in two-fold F. However, the metastable structure in Fig.~\hyperref[fig4]{4(d)} features edge-sharing and singly-coordinated F atoms. In spite of the significant difference in the short-range order, $\Delta$\textit{E}$_\textup{tot}$ is only 11.5 meV/atom. This suggests that quaternary InTe$_2$O$_5$F has a rugged energy landscape where the configurational similarity weakly correlates with the energy.  

\changemarker{In comparison with the DFT-based method, we find that some aspects of the evolutionary algorithm work differently because of characteristics of NNP. For instance, random and mutated structures often relax to unphysical structures by NNP because they possess short bonds that are not in the training set. In contrast, the DFT method can automatically adjust bond lengths. We believe that additional short-range repulsive potentials can relax those structures to reasonable ones that can be utilized by the evolutionary algorithm.}

\begin{figure}
\centering
 \noindent\makebox[\linewidth]{\includegraphics[scale=0.8]{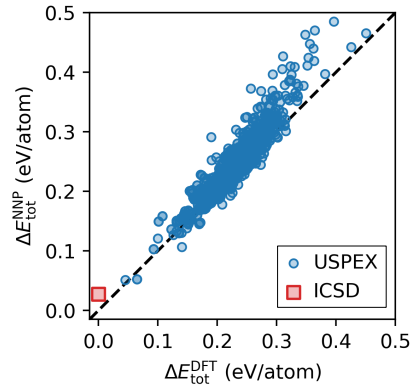}}
\caption{(Color online)  $\Delta$\textit{E}$_\textup{tot}^\textup{DFT}$ vs. $\Delta$\textit{E}$_\textup{tot}^\textup{NNP}$ for Ba$_2$AgSi$_3$ with structures in Fig.~\hyperref[fig2]{2(a)} after augmenting the training set with metastable structures from USPEX-NNP.  \label{fig5}}
\end{figure}

The metastable structures obtained from the evolutionary algorithm coupled with NNPs can be also used for updating NNPs. For instance, we add 22 candidate structures of Ba$_2$AgSi$_3$ to the training set and generate another NNP. [We recall that these are not structures in Fig.~\hyperref[fig2]{2(a)}.] In detail, we include the relaxation trajectories when DFT is applied on the candidate structures and energy curves along elastic deformations (shear, uniaxial, and hydrostatic types). After training, we recalculate NNP energies for metastable structures used in Fig.~\hyperref[fig2]{2(a)}. The result in Fig.~\ref{fig5} shows that energy resolution for the low-energy region is significantly improved in comparison with Fig.~\hyperref[fig2]{2(a)}, which is a result of hexagonal Si-Ag rings added to the training set. The Pearson coefficient for low-energy structures increases from 0.769 to 0.828. Notably, $\Delta$\textit{E}$_\textup{tot}$ for the third lowest metastable structure is 104 meV/atom in NNP, which is comparable to 93 meV/atom in DFT. Such refinement in the energy resolution should help identify the stable phase more quickly.  \changemarker{One can also improve NNPs through iterative training. For example, we perform a melt-quench simulation on Ba$_2$AgSi$_3$ using the NNP and re-train NNP by adding trajectories of amorphous phases to the training set. We then find that the energy correlation improves similarly to  Fig.~\ref{fig5}.}

So far, we assumed that \textit{Z} (the number of formula units in the unit cell) is known for the target compounds. This is not the case in dealing with a truly unknown compound and various \textit{Z} values should be tested separately. (We find that \textit{Z} for most inorganic crystals are 1, 2, or 4.) Utilizing NNP is advantageous in increasing the \textit{Z} number because of order-\textit{N} scaling, which is much more favorable than order-\textit{N}$^3$ scaling with DFT.

\subsection{Computational efficiency}
We remark on the computational efficiency of the present approach. The main computational cost of this work came from generating the training set, which requires tens of thousands of supercell calculations for MD. However, during MD simulations, wave functions extrapolate well, which significantly shortens self-consistent loops. \changemarker{As a computational hardware, we employed clusters of Intel\textsuperscript{\textregistered} Xeon Phi 7250 1.4 GHz (68 cores per CPU). CPU times spent over one generation are 2,072 (69,564) and 11,532 (244,646) secs/CPU for LiAlCl$_4$ and InTe$_2$O$_5$F, respectively, when NNP (DFT) is used. Using 4(8)-CPU cluster computers for ternary (quaternary) compounds, it took a couple of days in computer time} to obtain the whole training set and half-day for training and one day for USPEX to complete 120 generations with the NNP. With the same computational resource and the total time span (3--4 days), DFT-based USPEX could finish about 10-20 generations. 

As mentioned above, the multinary CSP would require generations significantly longer than 100. For instance, in evolving over 1,000 generations, NNP-based USPEX would be faster than DFT-based one by \changemarker{up to 30 times}. Furthermore, \changemarker{in evaluation of the energy and force of a single structure in the present study, NNP takes $1\times10^{-4}\sim3\times10^{-4}$ CPU seconds per atom while DFT takes $1\times10^{-1}\sim5\times10^{-1}$ CPU seconds per atom. Thus, the $\sim10^3$ speed gain by NNP did not translate into the acceleration in the present structure search by USPEX. This could be improved by tuning the parameters in USPEX and adjusting program interfacing between USPEX and LAMMPS.} By optimizing the supercell size and melt-quench protocols, it would be also possible to reduce the computational time to construct the training set significantly.

\section{\label{sec4}SUMMARY AND CONCLUSIONS}
In summary, we proposed a way to train NNPs using disordered structures sampled from liquid-quench-annealing MD trajectories. From the strong correlations of NNP and DFT energies among diverse metastable structures, it was confirmed that the NNPs can properly rank energies of structures that emerge from search algorithms. The CSP using the NNPs could identify low-energy metastable crystal structures more efficiently compared to the DFT-based approach. The short-range orders were established rather quickly but it seems to take much longer generations to arrive at correct medium-range orders. The fast energy evaluation by NNP would be advantageous for identifying medium range orders of multinary crystals from extended generations. \changemarker{We note that the present approach is equally applicable to unary and binary systems although DFT-based search methods can identify equilibrium structures efficiently for low-order compounds, as shown in numerous literatures.} In conclusion, by proposing a way to develop machine-learning potentials for CSP, this work will pave the way to identifying unexplored multinary phases efficiently.

\begin{acknowledgments}
This work was supported by Korea Institute of Science and Technology Information (KISTI) and Creative Materials Discovery Program through the National Research Foundation of Korea (2017M3D1A1040688). The computation was carried out at the KISTI National Supercomputing Center (KSC-2020-CRE-0064).
\end{acknowledgments}

\bibliographystyle{apsrev4-2}
\bibliography{manuscript}

\end{document}